# Field-free switching of perpendicular magnetization in an ultrathin epitaxial magnetic insulator

Sajid Husain[1*], Olivier Fayet[1], Nicholas F. Prestes[1], Sophie Collin[1], Florian Godel[1], Eric Jacquet[1], Thibaud Denneulin[2], Rafal E. Dunin-Borkowski[2], André Thiaville[3], Manuel Bibes[1], Nicolas Reyren[1], Henri Jaffrès[1], Albert Fert[1], and Jean-Marie George[1]

[1]Unité Mixte de Physique, CNRS, Thales, Université Paris-Saclay, 91767 Palaiseau, France.

[2]Ernst Ruska-Centre for Microscopy and Spectroscopy with Electrons, Forschungszentrum Jülich, 52425 Jülich, Germany.

[3]Laboratoire de Physique des Solides, Université Paris-Saclay, CNRS, 91405, Orsay, France.

*Present address: Material Physics Division, Lawrence Berkeley National Laboratory, Berkeley, California 94720, USA

jeanmarie.george@cnrs-thales.fr

albert.fert@cnrs-thales.fr

shusain@lbl.gov

**For energy efficient and fast magnetic memories, switching of perpendicular magnetization by the spin-orbit torque (SOT) appears as a very promising solution, even more using magnetic insulators that suppress electrical shunting. This SOT switching generally requires the assistance of an in-plane magnetic field to break the symmetry. Here, we present experiments demonstrating the field-free SOT switching of perpendicularly magnetized layers of the thulium iron garnet ($Tm_3Fe_5O_{12}$) magnetic insulator. The polarity of the switching loops, clockwise (CW) or counter-clockwise (CCW), is determined by the direction of the initial current pulses, in contrast with field-assisted switchings in which this polarity is controlled by the direction of the field. After an independent determination of Dzyaloshinskii-Moriya interaction (DMI), we relate the field free switching to the interplay of SOT and DMI and the polarity of the loops to the imprint of a Néel domain wall induced by the first pulse, in agreement with Kerr imaging. Our observation and interpretation of field-free electrical switching of a magnetic insulator is an important milestone for future low power spintronic devices.**

The pure electrical control of the magnetization in magnetic heterostructure employing spin-orbit torques (SOTs) is proposed as the key for developing energy efficient next generation magnetic memories in spintronics[1,2,3]. The manipulation of the magnetization through SOTs is usually realized using the spin current provided by the large spin-orbit coupling (SOC) found in heavy metals (HM). The spin current appears due to asymmetric spin scatterings mediated by SOC within the HM via the spin Hall effect (SHE) and/or the Rashba-Edelstein effect (REE) originating from interfacial SOC or an E-field orthogonal to the film surface (z-direction). The spin relaxation in the magnetic layer gives rise to a torque on the magnetization ($\mathbf{M} = M_s \cdot \mathbf{m}$), only the spins transverse to **m** relax and contribute to the torques. Mainly two torques have been identified as represented by a transverse (field-like 'FL') torque, $\tau_{FL} \propto \mathbf{m} \times \boldsymbol{\sigma}$ and a longitudinal (damping-like 'DL') torque, $\tau_{DL} \propto \mathbf{m} \times (\mathbf{m} \times \boldsymbol{\sigma})$, where $\boldsymbol{\sigma}$ is directed along the y-direction orthogonal to the spin current $J_s$. The out-of-plane torque due to structural symmetry-breaking is not considered here due to cubic symmetry of Pt. The driving force for the magnetization reversal in a perpendicular magnet is essentially the DL torque, but the FL can influence the nucleation[4,5]. Additionally, an in-plane magnetic field is commonly required for breaking the inversion symmetry[6]. It will favor one of the two magnetic remnant states, for a uniform magnetization switching (macrospin model[7]), and for domain nucleation and propagation mechanism as well[8]. A large effort has been dedicated to achieving field-free switching, as it would be an important milestone for low-power energy application.

Field-free switching has been recently studied in several systems, but this typically requires an additional process such as breaking crystal symmetry[9] to induce an additional torque, develop a chiral field gradient[10], etc. This usually invokes the fabrication of engineered multilayer stacks exhibiting a source of symmetry breaking. Moreover, in order to avoid the shunting of electrical current through metallic FMs, it is of interest to replace the metallic magnetic electrodes by a low conductivity material. On the other hand, ferrimagnetic (FIM) insulating garnets have recently entered into field of spin-orbitronics[11] opening new questions and issues about the nature of interfacial electronic exchange mechanisms when using a ferromagnetic insulator with a SHE metallic material. Recently, thulium iron garnet ($Tm_3Fe_5O_{12}$, TmIG) became one of the most attractive garnets due to its compelling properties such as perpendicular magnetic anisotropy (PMA) at room temperature[11], magnetization switching by SOT[12,13], large domain wall velocity[14], and interfacial chiral exchange, i.e., Dzyaloshinskii-Moriya interaction (DMI) as claimed in Ref.[15]. Several key features such as PMA strength and domain velocity are tunable through substrate-film strain engineering as well as through thin film growth[16,17,18]. The DMI has emerged as a debatable parameter of different possible origin. For example, it was reported to arise from the interface with a HM[19], from the substrate-film interface as well as from the bulk of the film[16,20]. Further, it is reported that chiral domains can be stabilized by DMI and moved by current pulses without external field[18]. A detailed analysis of DMI as well as its influence on the current induced switching in TmIG insulating garnet is still pending. TmIG grown along [1 1 1] also display cubic magnetocrystalline anisotropy[21], expected to provide an additional source for magnetization switching[22,23].

In this letter, we demonstrate the fabrication of high crystalline quality TmIG with perpendicular magnetization by off-axis sputtering. The major result of our study is the demonstration of the magnetization reversal induced by SOTs at zero field. We then discuss the field-free magnetization switching in light of Kerr microscopy imaging. We demonstrate the key role of the DMI in this system favoring a deterministic domain nucleation[5,10] depending on the polarity of the current and we will discuss the role of the magnetocrystalline anisotropy in the zero-field switching. To do this, we grew series of samples of TmIG on $Gd_3Ga_5O_{12}$ (GGG) substrates and Pt is used as a spin current source. See Methods for more description of the sample fabrication and measurement techniques.

We first perform a detailed characterization of the film structural quality, as DMI and PMA might find their origin in growth-induced strain and gradients. The 2θ – ω diffraction pattern of TmIG (15 nm) thin film deposited on GGG is displayed in Figure 1a. The corresponding peaks are identified as (4 4 4) reflections from both the TmIG and GGG substrate. Due to the small lattice mismatch (-0.49%) between the substrate and film (both are cubic crystals), the (4 4 4) diffraction peaks partly overlapped. The Laue oscillations visible around the main peak correspond to the film thickness fringes, indicating a high-quality growth and well-defined crystallographic ordering. The out-of-plane lattice spacing for the (4 4 4) reflection, $d_{TmIG}^{444}$ = 0.1775 nm, corresponds to the out-of-plane lattice constant $a_\perp$ = 1.230 nm. To confirm the epitaxial quality, we recorded a rocking curve along the (4 4 4) with a full width and half maximum Δω = 0.02° (compared to the 0.01° found for the GGG substrate) suggesting an epitaxial growth with a few millidegrees mosaicity. The RHEED pattern of TmIG after annealing displays a single family of streaks, indicating epitaxy and in-plane crystal coherence (Figure 1b). We measured the topography of the film surfaces using atomic force microscopy (AFM) (Figure 1c), and found a surface roughness smaller than 3Å. Furthermore, the reciprocal space mapping (RSM) was performed near the (6 4 2) asymmetric reflection, showing a pseudomorphic growth (Fig. 1d). The film peak and substrate peak at the same in-plane reciprocal lattice unit $d^{220}_{TmIG}$ = $Q_x^{-1}$ = 0.4379 nm, indicating a fully strained state with an in-plane lattice constant of 1.238 nm. Due to the elastic deformation of the TmIG under strain, the out-of-plane lattice spacing in the TmIG film corresponding to the (4 4 4) in the RSM is $d^{444}_{TmIG}$ = $Q_z^{-1}$= 0.1770 nm, consistent with the XRD evaluated lattice spacing. Using the in-and out-of-plane lattice parameters, the angle ($\Theta$) between for the facets of the TmIG unit cell[24] is calculated to be 90.74°. This tilt (in top edge of the cubic crystal cell) appears due to the tensile-strain from the GGG substrate. Since the 15nm thick film is fully strained, we assume that all the thinner films are also strained (see Supplementary Information, section I). The stoichiometry was verified using x-ray photoelectrons spectroscopy (see Supplementary Information, section II). Further, in the cross-sectional STEM image (Figure 1e), the substrate film interface is not distinguishable due to the epitaxial alignment of the lattices and two materials show a similar Z-contrast. However, the interface can be resolved through chemical analysis as marked by the dotted line. The atomically resolved STEM image shown in the inset reveals a highly ordered TmIG grown on GGG.

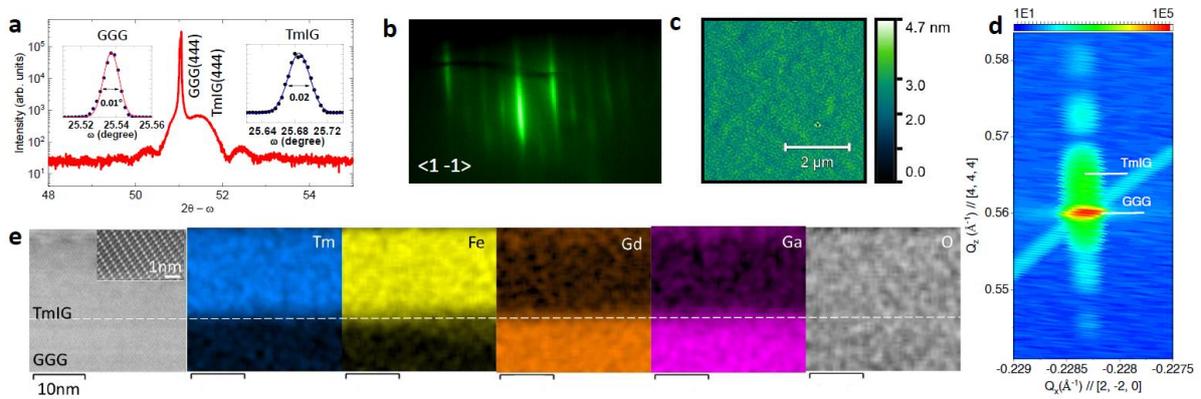

FIG. 1. **Structural characterization of TmIG. a** XRD pattern of TmIG(15 nm) grown on GGG. Inset on left (right) rocking curve of GGG (TmIG) of the (4 4 4) peak. **b** RHEED pattern of TmIG along the <1 -1> direction. **c** AFM topography of TmIG. **d** Reciprocal lattice mapping of TmIG along the (6 4 2) plane, which allow in-plane lattice measurement. Film and substrate reflections are marked. **e** Cross-sectional STEM imaging of TmIG/GGG along with the chemical assessment through electron dispersive x-ray analysis (EDX). Inset represents a closer look at the atom columns in the TmIG. The dotted line is drawn for TmIG/GGG interface. Scale bar, 10 nm.

Having confirmed that our TmIG films are of very good quality, we now discuss the SOT-induced switching of the TmIG perpendicular magnetization in several set of samples with different thicknesses of TmIG and Pt. As the interfacial DMI already observed in TmIG[18,19,20], appears to be involved in the switching mechanism of our samples, we have also measured the effective DMI using Brillouin Light Scattering (BLS). Depending on the sample, we find DMI energy densities ranging between 3.3 and 7.4 µJ/m² (corresponding to a DMI effective field applied to the domain walls between 1.8 to 4 mT), in the same range as the interfacial DMI found with TmIG in other publications[18,19,20]. All our quantitative BLS results are presented in Supplementary Information, section VIII.

In our switching experiments, charge current pulses (maximum pulse magnitude $J_c$ = 3 × 10¹¹ A/m²) are injected along the x-direction in the SOC material, here Pt. The SHE of Pt creates a spin current ($J_s$), which is injected into TmIG and propagates by magnons to generate DL and FT torques on its magnetization, involving DL and FL effective fields, **H**$_{DL}$ ∝ **m** × **σ** and **H**$_{FL}$ ∝ **σ** (see Fig.2a). The measurements of DL and FL fields are presented in the Supplementary Information, section VII. To demonstrate the effect of these torques on the magnetization, we perform fully reversible magnetization switching in a Hall bar using current pulses only. The magnetization is measured by anomalous Hall effect (AHE), and we obtain magnetization loops as a function of the pulsed current (Figure 2b). The initial magnetization state is first prepared by applying a magnetic field of 0.35 T in the out-of-plane direction and then reducing it to zero to obtain the initial state (state A with negative $R_{AHE}$ in Fig.2b). Then, still at zero field, we send a sequence of current pulses of magnitude $|J_c|$ up to $3.5 \times 10^{11}$ A/m². After each electrical pulse (of 100 µs width), a small reading current of (100µA = $6.7 \times 10^9$ A/m²) is used to detect the magnetization from the acquired values of the AHE.

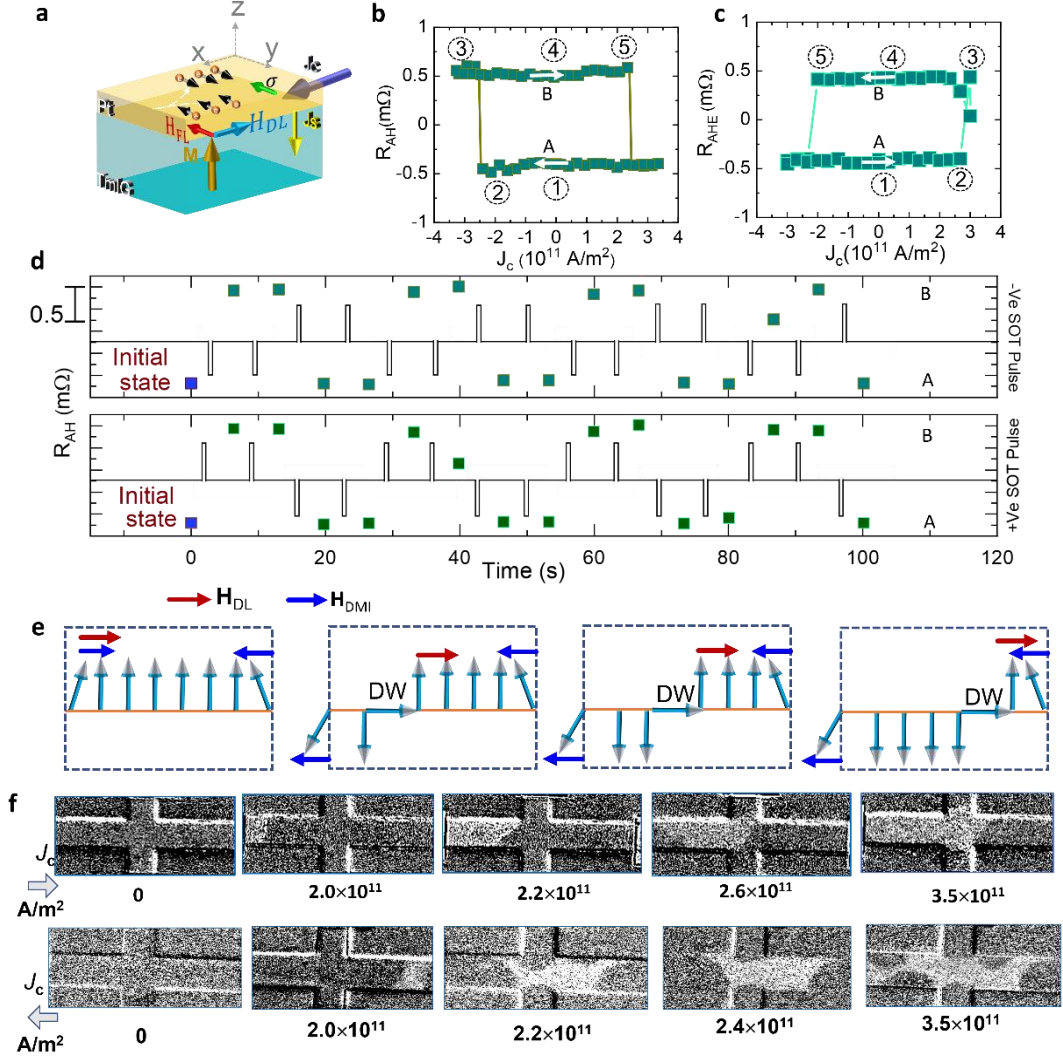

FIG 2. **Field-free switching in TmIG/Pt. a** Schematic of the current induced SOT fields on TmIG. **b,c,d, e** The sequence of current pulses shown in top (bottom) of **d** induces the type of CW (CCW) switching loops shown in **b** (**c**) : in **b**, starting from initial state A, a negative current pulse can switch TmIG from negative AHE (A) to positive AHE (B) and only a positive pulse (third in the sequence) can switch it back to A, with successive Clockwise (CW) loops generated by alternating negative and positive pulses. In **c**, a succession of CCW loops is obtained from the same initial state A but with an initial positive current pulse. The decider between CW (**b**) and CCW (**c**) is the sign of the first pulse. A mechanism consistent with these observations is pictured in **e** with current flowing from right to left (cf. pulse at point 2 in (**b**)), DMI at left edge of device helping SOT to start switching and DMI at right edge leading to the remanence of a small non-reversed domain. An opposite pulse (at point 5 in (**b**)) can expend this non-reversed domain to obtain the CW loop for the AHE (explained in the text). The blue (red) arrows indicate the direction of the DMI-induced fields '$H_{DMI}$' (DL fields, toward right for negative pulse). A positive (negative) first pulse leads to a remanent DW at left (right) and to CCW (CW) loops. The FL torque is not considered here. **f** Kerr imaging of the reversal by positive (negative) initial current pulses in top (bottom) panel at zero field, with approximated features of reversal starting on the left (right) edge and propagating to the right (left) with more complex behavior at the crossing of the Hall contacts. Colors black and white represent as magnetization up and down, respectively. Images were recorded by initializing the state with 20 mT before each higher current pulse.

In the pulse sequence starting with a negative pulse in the schematic of the top panel in Figure 2d, when the negative pulse exceeds a critical value, $R_{AHE}$ switches abruptly from negative to positive (from (2) to (3)) and goes to state B at the end of the pulse in Fig.2b, (switch of $m_z$ from positive to negative). In the continuation of the sequence in the top panel in Figure 2d, switching back to $R_{AHE}<0$ can be obtained only by positive pulses, as in the succession of the experimental CW loops Fig.2b. We note that the switching ratio $R_{SOT}^{Pulse}/R_{AHE}$ = 0.52/0.55 is found to be ~95%, which means almost complete switching ($R_{AHE}$ is taken from the AHE measurements, see Supplementary Information, section VI). The results are quite different with the second type of pulse sequence (bottom panel in Fig. 2d) starting from the same initial state (A) with first positive pulses. An abrupt switching from A to B (from positive to negative $m_z$) is now obtained with the first positive pulse, from (2) to (3) in Fig.2d, and the system comes back to A with negative pulse, from (5) to (1) in the CCW loop in Fig.2c. With the first sweep of positive current pulses, CCW loops replace the CW loops obtained with a negative first pulse in Fig.2b. Thus, after the same magnetic initialization, the system behaves differently depending on its first current-induced switching pulse.

We recall the usual situation in which the addition of an applied in-plane field along x is needed to break the in-plane symmetry and switch a perpendicular magnetization by SOT[3,4]. In such well-known experimental protocol, the direction of this field along x decides that the switching loop is CW or CCW[3,4]. In our switching results free of any external field, the decider of the choice between CW and CCW is the direction of the initial current pulses. Such a behavior has already been found[10] and is ascribed to the chiral remanence imprinted by the first pulse in systems with a DMI-induced field $H_{DMI}$ tilting the spins at the edges[25]. Figure 2e describes an example of mechanism of this type, which involves $H_{DL}$ and DMI and is consistent with our results. For a given direction of the current pulse and the corresponding direction of $H_{DL}$ (red arrow) in Fig.2e, the switching to down starts on the left edge where $H_{DMI}$ (blue arrow) helps $H_{DL}$ (red arrow) and propagates to right by Néel DW motion. This motion and the corresponding switching to down stops at the approach of the opposite edge where $H_{DMI}$ (blue) hinders $H_{DL}$ (red), what imprints a remanent non-switched up domain close to the edge (a chiral remanence). An opposite pulse reverses $H_{DL}$ and enlarges the remanent up domain to the right. It gives a CCW loop for $m_z$ (CW for $R_{AHE}$) in this case. With a first pulse of opposite amplitude, the switching starts at the left edge and it is easy to see that it leads to a CW loop (CCW for $R_{AHE}$) of opposite polarities. Adding $H_{FL}$, crystal field or more complicated shape of the device leads to some variants of this type of mechanism, as discussed later. In samples with defects, we could also consider the possibility of chiral remanences on defects in the bulk of the layer[26].

Kerr imaging (Fig.2f) shows that the experimental behavior is similar close to the scenario in Fig.2e with, for the direction of the current pulse of the top (bottom), a reversal starting on the left (right) of the device (Fig.2f, top panel) and propagating to the right (left) with, however, a somewhat complex behavior when the domain wall arrives in the wider region of the Hall-cross. These images are recorded in zero in-plane applied magnetic field (only in earth's magnetic field). See Fig. S10 to rule any effect of spurious magnetic fields present in the measurement setup.

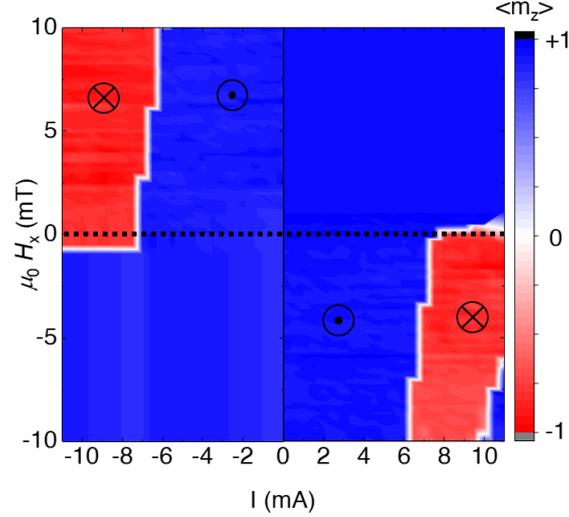

FIG. 3. **Current-field map of magnetization switching.** Experimental magnetization switching map recorded in the in-plane magnetic fields (along or opposite to the current direction) after initializing the magnetization with +0.35 T magnetic field. Dotted line is drawn at zero-field crossover.

In order to fully characterize the magnetization reversal process, we also discuss the influence of a magnetic field, as reported in Fig.3. In zero field, starting from an initial state with magnetization up ($R_{AHE} < 0$, point A in Fig.2b), the system switches from up to down (blue to red) at a negative current of about ~7 mA (~2.5 $10^{11}$ A/m$^2$ in Fig.2b). The application of a positive field $\mu_0 H_x$ helps to switch and the switching current decreases in absolute value down to about -6 mA at ~10 mT. In contrast, a negative field appears to hinder the switching by a negative current and suppresses it above about -0.9 mT. For positive fields, the switching from up to down (blue to red) occurs in positive currents, which correspond to a change of the polarity of the loops, from the CW type of Fig.2b to the CCW if Fig.2c. The decrease of the switching current as the field increases expresses that negative fields help the switching of this polarity. We thus find that, in addition to the results obtained at zero field, an applied field can help or hinder the switching and even change the polarity of the loops. It is interesting to note that the field changing the polarity (-0.9 mT) for negative current is close to the DMI field derived from our BLS measurements (1.8 mT) and that the fields efficient to change the switching currents[27] are also in the same range. We also observed, depending on the Hall bar as presented in Supplementary Information, section XI, some asymmetry in the switching current polarity we relate to extrinsic mechanisms linked to the nucleation process (defect, shape, inhomogeneity etc..). A perfect control of the shape and the edges of the ferromagnetic insulator may consider for further development.

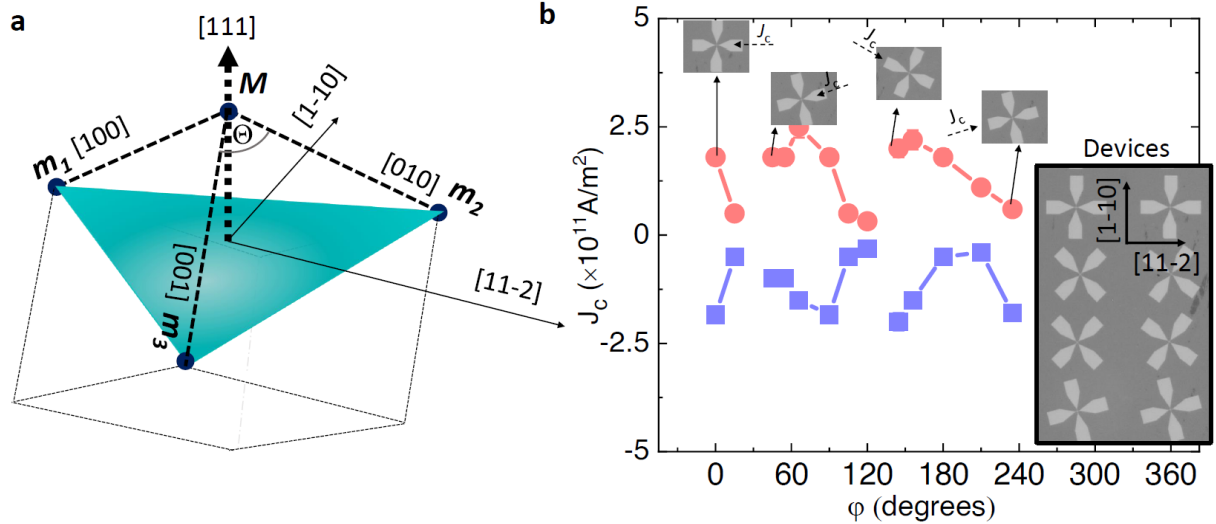

FIG. 4. **Role of magnetocrystalline anisotropy. a** Schematic of the [111] orientation of the TmIG crystal lattice with three vectors [1 0 0] [0 1 0] [0 0 1] of the magnetocrystalline anisotropy. The [1 1 -2] and [1 -1 0] are two in-plane direction vectors (as also depicted for substrate plane). The Θ is the angle of TmIG cubic crystal under strain from substrate. **b,** Patterned devices with different azimuthal angles. Critical current density plot for various devices patterned at different azimuthal angles for a positive $m_z$ initialization.

Although the field free switching we observed with a loop polarity controlled by the sign of the first current pulse in Fig.2 can be consistently explained by the conjunction of SOT and DMI in the mechanism summarized in Fig.2e, additional results indicate that the cubic anisotropy (see orientation of crystal axes at (111) surface in Fig.4) has also some influences onto the switching process. To understand the role of the cubic anisotropy experimentally, we patterned devices with different azimuthal angles as shown in inset of Figure 4b. The critical switching current recorded in these devices is found to be dependent on the crystallographic orientation, while the anomalous Hall effect signals were found to be identical with the same coercivity (not shown here), ruling out any non-uniformity in the samples. The orientation dependence of the critical current indicates the additional influence of the cubic anisotropy. We also measured the magnetization switching by varying the TmIG and Pt thicknesses (see Supplementary Information, section X): some variation in the anomalous Hall signal can be seen however field free switching is always observed. Further, micromagnetic simulations were performed at different combination of parameters demonstrating the critical role of DMI and cubic anisotropy in the magnetization reversal process by SOT (see Supplementary Information, section XII).

In summary, we have observed the magnetization switching of perpendicularly magnetized epitaxial $Tm_3Fe_5O_{12}$ (TmIG) thin films in TmIG/Pt bilayers for which we have the advantage of the deep propagation of spin currents by magnons for efficient SOT and the absence of shunting in TmIG. Another advantage is the existence of interfacial DMI that we determined by Brillouin Light Scattering. We observe a reproducible and robust field-free switching at moderate current density. Starting from a saturated magnetic state, the sign of the first switching current pulse decides if the subsequent switching loops are CW or CCW, in the same way as, in in-plane-field-assisted switching, the "decision" is taken by the sign of the external applied field. The main features of the field free switching results are consistent with a mechanism in which the conjunction of efficient SOT (DL) and DMI nucleates reversed domain on one edge of the sample and imprints a chiral remanence on the opposite edge. Kerr imaging is also consistent with such a mechanism. In experiments in which a field is applied, we find that a field in the range of the DMI fields helps or hinders the switching observed at zero field and can even inverse the polarity of the loops. Additional experiments show that the cubic anisotropy plays an intriguing role in the switching at zero-field. While there remain outstanding questions as to the exact origin of field-free switching in TmIG, one plausible explanation we propose here is the DW nucleation at the edges due to the DMI and cubic anisotropy, which follows the current pulse direction. Future experimental and theoretical work should seek to further understand the role of different parameters of TmIG, such as different crystal orientations. In conclusion, the finding of field-free magnetization switching by SOTs in a magnetic insulator is an important milestone for future applications in spin-orbitronics.

**Acknowledgements**

DARPA TEE program grant (MIPR#HR0011831554) is acknowledged for their financial support. This work is supported by a public grant overseen by the French National Research Agency (ANR) as part of the "Investissements d'Avenir" program (Labex NanoSaclay, reference: ANR-10-LABX-0035). ERC AdG FRESCO (#833973) is also acknowledged.

# Methods

**Thin film growth:** Thulium iron garnet 'Tm$_3$Fe$_5$O$_{12}$ (TmIG)' ferrimagnetic insulator (FIMI) thin films were deposited on (1 1 1) oriented Gd$_3$Ga$_5$O$_{12}$ (GGG) substrates by off-axis sputtering. Before deposition, substrates were treated by acetone and isopropyl alcohol in ultrasonication and subsequently annealed at 1000°C for 5 hours in a flow of pure oxygen (O$_2$) at atmospheric pressure. The substrates were transferred in air into the sputtering chamber for TmIG deposition. Thin films were deposited at room temperature in flow of Ar (40 sccm) and O$_2$ (20 sccm) with dynamic pressure of 4.2 mbar (base pressure is lower than ~2×10$^{-8}$ mbar). To promote the crystallinity, these films were post-annealed (in ex-situ furnace) at 650 °C for 4 hours in a flow of pure O$_2$ at atmospheric pressure. Further, Pt layer of 6nm thick (unless otherwise stated) was deposited by on-axis magnetron sputtering at room temperature. TmIG film surfaces were cleaned by O$_2$ plasma (~40 eV) before Pt deposition.

**Structural characterization:** X-Ray diffraction in symmetric (2θ–ω) or asymmetric (reciprocal space mapping, RSM) geometry were recorded by Philips X'pert-PRO Empyrean diffractometer. For XRD, measurements were performed in Bragg-Brentano reflection mode. For RLM, the diffraction along the (6 4 2) plane direction is used, which allows to gain the information about in-plane epitaxy relation along [2 -2 0] direction. Topography of the substrate and film were recorded by atomic force microscopy (AFM) using a Dimension Icon system with ScanAsyst(Bruker Dimension Icon, Billerica, MA, USA). Images were collected in tapping mode (in air) using a tip with nominal radius <10 nm. Atomic-scale imaging was performed by cross-sectional scanning transmission electron microscopy (STEM). The sample investigated by STEM was prepared by a focused ion beam machine (FEI Helios platform) using a Ga ion beam with an accelerating voltage of first 30 kV to detach the slab, and then of 5 kV to thin it down. STEM characterization was conducted with a Hitachi HF5000 equipped with a cold field emission gun operated at 200 kV and a probe aberration corrector. High-angle annular dark-field images were acquired with a probe that formed an angle of 30 mrad and a collection angle of 60–300 mrad. EDX spectra were collected using two detectors from Oxford Instruments and color-coded elemental maps were obtained using the AZtec software. Magnetization measurements were performed by Quantum Design SQUID magnetometer. All electron transport measurements were performed in a home-built set-up.

**AHE, SMR, SOTs and magnetization switching measurements:** To perform the electron transport experiments, 5-μm wide and 50-μm long symmetric Hall-crosses were patterned using photo-lithography and Ar-ion milling. For Kerr imaging, the devices were patterned in 10-μm wide and 100-μm long Hall crosses with Au contact pads. The anomalous Hall effect and spin Hall magnetoresistance measurements were carried out using a constant dc source. For current-induced switching measurements, current pulses with a duration of 100 μs were generated by a Keithley 6221 and injected into the Hall bar. After each pulse, a small excitation (100 μA = $3\times10^9$ A/m$^2$) current was applied to evaluate and measure the magnetization state. For the harmonic Hall measurements, an ac current source with an amplitude from 1 to 6 mA (root mean square) was injected with a Keithley 6221 current source. The first and second harmonic signals were measured using an SR-830 lock-in amplifier.

**DMI measurements:** The Dzyaloshisnkii-Moriya interaction energy was measured by Brillouin light scattering (BLS) using a JRS TFP-2 triple-pass tandem Fabry-Perot interferometer with quarter-wave antireflection optics and linearly-polarized (10 mW laser with 473 nm wavelength). The spectra were recorded in the backscattering geometry at various wave vector orientations, selected by mounting the sample on an angle-controlled sample holder providing a range of 10° to 60° incident angles corresponding to wave vectors, $q_k = (4\pi/\lambda)\sin\theta$, lying in the range 4 to 20.4 rad./μm$^{-1}$. The free spectral range was 9.4 or 18.7 GHz and spectra were recorded with 1024 points. The in-plane magnetic field to pull the magnetization in-plane for the Damon-Eshbach geometry is provided by permanent magnets in order to avoid thermal drift.